\documentclass[10pt,conference]{IEEEtran}
\usepackage{cite}
\usepackage{amsmath,amssymb}
\usepackage{graphicx}
\usepackage{booktabs}
\usepackage{array,multirow}
\usepackage[table]{xcolor}
\usepackage{threeparttable}
\usepackage{microtype}
\usepackage{xurl}
\usepackage[hidelinks]{hyperref}
\usepackage{newtxtext,newtxmath}
\graphicspath{{figures/}}
\usepackage[most]{tcolorbox}
\definecolor{promptaccent}{RGB}{176,74,0}
\usepackage{tabularx}

\newif\ifanonymous
\anonymousfalse

\title{Merged, but Did It Fix the Issue?\\
Using LLMs to Link GitHub Issues to Their Fixes}
\ifanonymous
  \author{\IEEEauthorblockN{Anonymous Authors}}
\else
 \author{
\IEEEauthorblockN{Afridi Mahmud}
\IEEEauthorblockA{
Bangladesh University of\\
Engineering and Technology\\
Dhaka, Bangladesh\\
\texttt{afridimahmud99@gmail.com}
}
\and
\IEEEauthorblockN{Md. As-Aid Rahman}
\IEEEauthorblockA{
Bangladesh University of\\
Engineering and Technology\\
Dhaka, Bangladesh\\
\texttt{mdasaidrahman@gmail.com}
}
\and
\IEEEauthorblockN{Rifat Shahriyar}
\IEEEauthorblockA{
Bangladesh University of\\
Engineering and Technology\\
Dhaka, Bangladesh\\
\texttt{rifat@cse.buet.ac.bd}
}
}
\fi

\begin{document}
\maketitle
\begin{abstract}
A GitHub issue can link to several pull requests (PRs), and even a merged PR may not fix it. Benchmarks for coding agents need the PR that fixed each issue, or need to know that none did. We labeled 1,051 issues from 699 repositories and compared six non-LLM methods with eleven LLM setups. Closing keywords such as ``Fixes~\#N'' were precise but missed about a third of fixes. Picking any merged PR never recognized that no PR was the fix. The strongest LLM setups had similar overall accuracy (92--93\%) and found merged fixes at similar rates. They differed when PRs were merged but none fixed the issue: Claude Sonnet~5 answered ``no fix'' correctly on 88\% of these issues and Qwen3-4B on 53\%. Sonnet was significantly better than every model we ran locally, and changing the open models' cut-off did not reach it. Within one model family, this ability grew with model size. At the estimated rate of such issues, Sonnet would keep about a quarter fewer wrong reference fixes than Qwen3-4B, but reject more real ones. Most open models also over-applied a prompt exception needed for only 1\% of fixes. We release the labels, model outputs, and 6,785 model-produced labels.

\end{abstract}

\section{Introduction}\label{sec:introduction}
Consider a GitHub issue reporting that a configuration parser crashes on an empty file. Three pull requests (PRs) link to it. The first proposes a fix but is never merged. The second is merged but only adds tests. The third is merged and fixes the parser. Only the third is the fix, and sometimes none of the linked PRs is.

This happens in practice. In issue \#2617 of EleutherAI's \texttt{lm-evaluation-harness}, a merged PR says ``Fixes \#2617,'' but its author writes that it ``does not fix the slowness in itself.'' A rule that trusts closing keywords, or one that picks any merged PR, selects this PR as the fix.

Such mistakes matter for benchmarks of coding agents, which build tasks from GitHub issues and their linked code changes~\cite{jimenez2024swebench,pan2025swegym,badertdinov2025swerebench}. A wrong reference fix produces a task that does not match its issue, and the risk grows when pipelines select reference PRs automatically. Prior work has examined benchmark quality through tests and human review~\cite{openai2024swebenchverified,bhatia2025spice}, recovered changes related to an issue~\cite{sun2017frlink,huang2026easylink,zheng2026selink,akhavan2026linkanchor}, and checked whether a given issue--PR pair matches~\cite{isik2025prreviews,wang2026paichecker}. We study a different setting: picking the fix from several linked PRs, or deciding that none of them is the fix.

Finding the fix involves two questions: was the PR accepted, and did it fix the issue? Merge status answers the first question well, although fixes occasionally land without merging. The second question remains even among merged PRs, which may add tests, documentation, or related changes without fixing the reported problem.

We study this task using 1,051 manually labeled issues from 699 repositories, comparing six non-LLM methods with eleven LLM setups. We ask:
\begin{itemize}
    \item \textbf{RQ1:} How often do closing keywords point to the fix?
    \item \textbf{RQ2:} Do LLMs follow the rule that unmerged PRs are not fixes?
    \item \textbf{RQ3:} Can LLMs tell when a merged PR did not fix the issue?
\end{itemize}

Our contributions are:
\begin{itemize}
    \item \textbf{A labeled dataset} of 1,051 issues, including competing merged PRs and 311 issues with no fix, released with model outputs and 6,785 additional model-produced labels.
    \item \textbf{Evidence that simple signals fail in both directions, and that prompts are a weak way to enforce rules.} Closing keywords miss about a third of fixes, picking any merged PR never says ``no fix,'' and a rarely needed exception in the prompt made several models break the merge rule far more often.
    \item \textbf{Evidence that models differ most in noticing when a merged PR is not the fix.} At default settings, models with near-identical overall accuracy found merged fixes at similar rates but differed widely on this case, and the models that noticed it more often also missed more real fixes. Moving an open model's ``no fix'' threshold shifts it along this trade-off but does not reach the strongest model.
\end{itemize}

\section{Background and Related Work}\label{sec:related}
\subsection{Issue--Fix Link Recovery}

Early methods used issue text, commit messages, changed files, and other project records to recover missing links~\cite{wu2011relink,sun2017frlink,rostami2021hybridlinker}. DeepLink and EALink trained models to match issues with commits, and MPLinker used prompts~\cite{ruan2019deeplink,zhang2023ealink,wang2025mplinker}. EasyLink and SELink rank candidate commits with an LLM~\cite{huang2026easylink,zheng2026selink}, while LinkAnchor uses an LLM agent to search project history~\cite{akhavan2026linkanchor}. Morgan et al.\ found that trained rankers without an LLM can outperform LLM rankers~\cite{morgan2026rethinking}. These methods find changes related to an issue; they do not decide which of several related PRs is the fix.

\subsection{Issue--PR Alignment and Benchmark Quality}

Wrong bug--fix links can bias software engineering data~\cite{bird2009bias}, and many fixes are never linked~\cite{bachmann2010missing}. SWE-bench builds tasks from GitHub issues and their linked changes~\cite{jimenez2024swebench}, as do SWE-Gym, SWE-rebench, and SWE-bench-Live~\cite{pan2025swegym,badertdinov2025swerebench,zhang2025swebenchlive}. SWE-Bench+ identified tasks with leaked solutions or weak tests~\cite{xue2026swebenchplus}; SWE-bench Verified uses human review~\cite{openai2024swebenchverified}, and SPICE automatically labels issue clarity, test coverage, and effort~\cite{bhatia2025spice}. I\c{s}\i k et al.\ identified PRs whose changes differ from what the issue requested~\cite{isik2025prreviews}, and PAIChecker found mismatches in 13.6\% of SWE-bench Verified tasks by using LLM agents to assess one issue--PR pair at a time~\cite{wang2026paichecker}. We instead compare several linked PRs and allow the answer that none of them is the fix.

\subsection{Following Rules and Answering ``No Fix''}

Prior work has documented failures of LLMs to follow explicit instructions~\cite{zhou2023ifeval,jiang2024followbench} and to decline to answer when appropriate~\cite{wen2025abstention,kirichenko2025abstentionbench}. Our task involves related decisions: RQ2 examines whether models follow a merge rule, and RQ3 examines when they correctly or wrongly answer ``no fix.'' Unlike declining because of uncertainty, ``no fix'' is a valid answer when none of the linked PRs is the fix.

\section{Dataset}
\label{sec:dataset}
\subsection{Task}
\label{sec:task}
A method receives a GitHub issue and its linked PRs. It must select the \emph{main fix}, the one PR the project accepted as the fix for the issue, or answer ``no fix'' if none qualifies. In the rest of the paper we call it simply \emph{the fix}. We assign each linked PR one of three labels:
\begin{itemize}
    \item \textbf{Fix:} the accepted PR that fixed the issue.
    \item \textbf{Helper PR:} an accepted PR that made related changes, such as tests, documentation, or follow-up work, but is not the fix.
    \item \textbf{Not a fix:} a PR that was not accepted, or that is unrelated to the issue.
\end{itemize}
Acceptance usually means merging, but maintainers sometimes apply changes through separate commits. Our labels also count such PRs as accepted when the records establish that their changes landed (Section~\ref{sec:mergerate}).

\subsection{Collecting and Labeling Issues}
\label{sec:labeling}
We searched public, non-fork, non-archived GitHub repositories across languages and popularity levels, collecting 7,985 recently closed issues from 1,972 repositories, together with their same-repository PRs linked through references or timeline events. Simple rules assigned each issue a likely fix and a confidence level. We sampled across confidence levels and numbers of linked PRs, oversampling unclear cases and competing PRs to obtain enough difficult cases for analysis. About 35\% of labeled issues came from groups the rules marked unclear, compared with 5\% of unlabeled issues. Estimates for all mined issues account for this sampling (Sections~\ref{sec:rq1} and~\ref{sec:rq3}).

One author labeled each issue using saved issue discussions and PR records, without seeing model outputs or the rules' labels. The records included changed files and change size but not code diffs; we did not run code or tests. A second annotator independently labeled 200 issues (19\%, 510 PRs), with 95.5\% issue-level agreement (Cohen's $\kappa=0.92$) and 98.4\% PR-level agreement ($\kappa=0.975$). Both selected the same PR in all 110 issues where they identified a fix. Disagreements were resolved through discussion. Two additional annotators later labeled 100 test issues, 52 with merged PRs but no fix and 48 with a merged fix; we use their labels to check RQ3 (Section~\ref{sec:rq3}).

We excluded 149 issues outside the task: 66 tracking or umbrella issues, 50 duplicate or invalid issues, 23 with several equally important fixes, and 10 for other reasons. The resulting dataset contains 1,051 issues from 699 repositories (Table~\ref{tab:dataset}). Of these, 71 were used to refine the labeling guidelines. We split the remaining issues by repository into 305 development and 675 test issues, with no repository shared between the splits; the other 33 repositories contributed only guideline issues.

\begin{table}[t]
\centering
\caption{Dataset summary (totals include 71 guideline issues).}
\label{tab:dataset}
\small
\setlength{\tabcolsep}{5pt}
\renewcommand{\arraystretch}{1.05}
\begin{tabular*}{0.96\columnwidth}{@{\extracolsep{\fill}}lrrr@{}}
\toprule
& \textbf{Total} & \textbf{Dev} & \textbf{Test} \\
\midrule
Repositories & 699 & 218 & 448 \\
\textbf{Issues} & \textbf{1,051} & \textbf{305} & \textbf{675} \\
\quad With a fix & 740 & 213 & 482 \\
\quad No fix & 311 & 92 & 193 \\
\midrule
\textbf{Linked PRs} & \textbf{1,944} & \textbf{546} & \textbf{1,214} \\
\quad Fix & 740 & 213 & 482 \\
\quad Helper PR & 193 & 51 & 109 \\
\quad Not a fix & 1,011 & 282 & 623 \\
\bottomrule
\end{tabular*}
\end{table}

\subsection{Issue Groups and Merge Status}
\label{sec:groups}
\label{sec:mergerate}
\begin{figure}[t]
\centering
\resizebox{\columnwidth}{!}{%
\begin{tikzpicture}[
  font=\sffamily\scriptsize,
  box/.style={draw=black!60, fill=black!3, rounded corners=2pt, align=center,
              text width=1.6cm, minimum height=2.0cm, inner sep=2pt},
  arr/.style={->, >=stealth, semithick, black!60},
  rq/.style={font=\sffamily\tiny\bfseries, text=promptaccent, anchor=north}]
\node[box] (mine)  at (0.87,0) {\textbf{Mine}\\[3pt]7,985 issues\\[2pt]1,972 repos};
\node[box] (label) at (2.86,0) {\textbf{Label}\\[3pt]1,051 issues\\[2pt]dev 305\\[2pt]test 675};
\node[box] (pick)  at (4.85,0) {\textbf{Pick the fix}\\[3pt]6 non-LLM\\[2pt]11 LLM setups};
\node[box, text width=2.5cm] (rule) at (7.29,0) {\textbf{Merged-only}\\[3pt]\mbox{unmerged $\rightarrow$ no fix}};
\node[box, text width=2.2cm] (eval) at (10.03,0) {\textbf{Test groups}\\[3pt]no merged PR 139\\[2pt]merged fix 476\\[2pt]\textcolor{promptaccent}{\textbf{merged, no fix 60}}};
\draw[arr] (mine.east) -- (label.west);
\draw[arr] (label.east) -- (pick.west);
\draw[arr] (pick.east) -- (rule.west);
\draw[arr] (rule.east) -- (eval.west);
\node[rq] at ([yshift=-0.6mm]pick.south) {RQ1};
\node[rq] at ([yshift=-0.6mm]rule.south) {RQ2};
\node[rq] at ([yshift=-0.6mm]eval.south) {RQ3};
\end{tikzpicture}}
\caption{Study pipeline. RQ1 checks closing keywords, RQ2 checks rule-following before the merged-only rule, and RQ3 compares methods on test issues with a merged PR, focusing on the merged-no-fix group.}
\label{fig:pipeline}
\end{figure}
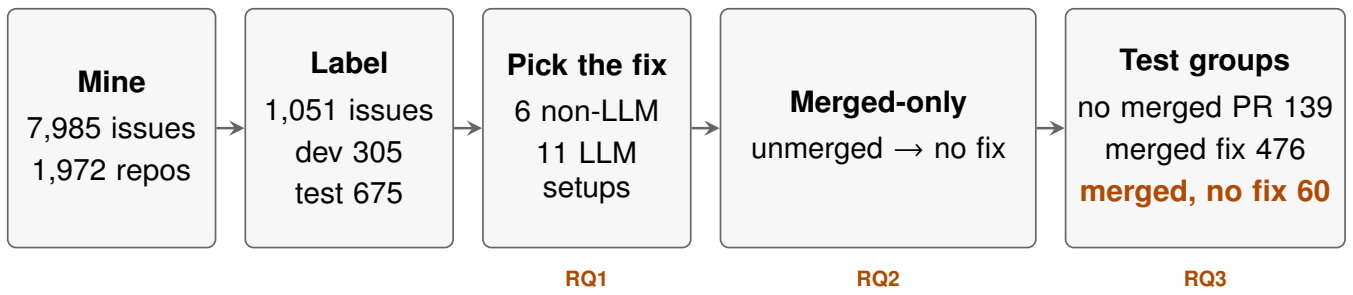

We divide the 675 test issues by their linked PRs (Fig.~\ref{fig:pipeline}): 139 have no merged PR; of the 536 with a merged PR, 476 have a merged fix and 60 have merged PRs but no fix. Of the 536, 147 have two or more merged PRs, spanning both groups. The development set has 31 merged-no-fix issues.

Of the 482 test issues with a fix, 476 (98.8\%) had a merged fix. The other six fixes were applied through separate commits, as confirmed by maintainer comments, and all belonged to issues with no merged PR. This high merge rate partly follows from our labeling criteria: an unmerged PR counts as accepted only when the records establish that its changes landed, so undocumented landings may be missed.

To look for such landings, we queried GitHub for all 707 unmerged development and test PRs labeled ``not a fix.'' We checked whether each PR was closed by a commit, whether its commits appeared on the default branch, and whether a default-branch commit referenced it. After excluding changes that landed through another merged candidate PR and references that did not indicate a landing, 20 PRs across 18 issues (14 test issues) remained possible silent landings. Even if all 14 test issues had additional unmerged fixes, 476 of 496 fixes (96.0\%) would still be merged.

Merged, however, does not mean fixed. In the test set, 109 merged PRs were helpers, and 60 issues had merged PRs but no fix.

\section{Approach and Experimental Setup}\label{sec:approach}
\subsection{Asking an LLM to Pick the Fix}
For each issue, an LLM selects the fix from the linked PRs or answers ``no fix'' (\texttt{null}). It receives the issue title, description, and comments, along with each PR's title, description, review discussion, merge status, changed files, change size, and commit information. These are the same records used for labeling; models have no repository or tool access. The base prompt (Fig.~\ref{fig:prompt}) distinguishes fixes from helper PRs and asks for evidence of acceptance. It allows an unmerged PR only if a maintainer states that its changes landed another way; the \emph{strict prompt} removes this exception. Models answer once per issue in JSON (chosen PR or \texttt{null}, helper PRs, and a short reason), without fine-tuning. Unreadable answers count as ``no fix''; they were rare (none for nine setups, one for Phi-4-mini, and four for Qwen3-4B in thinking mode).

\begin{figure}[t]
\centering
\providecolor{promptaccent}{RGB}{176,74,0}
\begin{tcolorbox}[
  width=\columnwidth,
  colback=white,
  colframe=black!60,
  colbacktitle=black!75,
  coltitle=white,
  fonttitle=\bfseries\sffamily\scriptsize,
  title={SYSTEM PROMPT \hfill \normalfont abridged},
  boxrule=0.5pt,
  arc=1mm,
  left=5pt, right=5pt, top=4pt, bottom=4pt,
  fontupper=\sffamily\footnotesize,
  fontlower=\sffamily\footnotesize
]
\raggedright
\textbf{Canonical solution:} the single PR that was
\textbf{accepted (merged or otherwise landed)}
and is the primary fix for the problem described in the issue.\\
\textbf{Auxiliary PR:} accepted, but only related, partial, follow-up,
docs/tests-only, or a revert/cleanup around the fix. Not canonical.\\
\textbf{No solution:} if no candidate PR was accepted as the fix, answer \texttt{null}.\\
\textbf{Evidence, strongest first:} GitHub closing/link events;
``fixes/closes/resolves \#N'' in the PR; a maintainer saying the PR fixed it;
merged PR whose change matches the issue; weak text similarity.\\
\textbf{A closed-unmerged PR is normally not a solution}
\textcolor{promptaccent}{\textbf{[unless a maintainer says it landed another way]}}.
\tcblower
\raggedright
{\scriptsize\bfseries\color{black!65}USER PROMPT: EACH CANDIDATE}\par\smallskip
\texttt{\#\#\# Candidate PR \#}\textit{n}\texttt{:} \textit{title}\\
\texttt{Status:} \texttt{MERGED at} \textit{time}
\enspace\textit{or}\enspace
\texttt{state=closed|open, NOT merged}\\
\textit{author, changed files, $+$/$-$ lines, commits; description; review discussion}
\end{tcolorbox}
\caption{Abridged base prompt used for the main model runs. ``Canonical'' denotes the fix and ``auxiliary'' a helper PR. The bracketed exception is removed in the strict prompt (Section~\ref{sec:rq2}). Full prompts are available in the artifacts.}
\label{fig:prompt}
\end{figure}

\subsection{Merged-Only Rule}
\label{sec:mergedonly}
Because recorded fixes were almost always merged (Section~\ref{sec:mergerate}), our main comparison considers only merged PRs. We apply this rule after the model answers, replacing picks of unmerged PRs with ``no fix,'' and keep the original answers to study rule-following in RQ2. Removing unmerged PRs from the input instead gives similar accuracy (91.7\% versus 92.1\% for Qwen3-4B).

The rule also excludes the six test fixes that landed without merging. After applying it, every method answers ``no fix'' on the 139 issues without a merged PR, correctly on 133 and wrongly on six. Since these issues no longer separate methods, RQ3 focuses on the 536 issues with at least one merged PR.

\subsection{Changing the Prompt and Input}
\label{sec:variants}
For RQ2, we change one prompt or input feature at a time and examine the original answers, before the merged-only rule. Prompt changes use a stricter acceptance rule, remove the exception for unmerged PRs, add an explicit merge flag, state that ``no fix'' is valid, or remove fix claims from the evidence list. Input changes add the issue timeline, commits, and code diffs; remove ``Fixes~\#N'' claims from unmerged PRs; hide or flip merge status; or remove unmerged PRs. We test these changes on Qwen3-4B, Qwen3-14B, Phi-4-mini, and Granite-3.3-8B, and test the strict prompt on all seven non-thinking open models and two hosted models.

We also examine which PR features go with picking an unmerged PR. For the 484 unmerged test PRs that are not the fix, we fit a logistic regression on features including closing keywords, whether the issue has a merged PR, whether the PR is open, its list position, and whether it has the largest change.

For RQ3, we also vary how readily an open model answers ``no fix.'' From Qwen3-4B's next-token probabilities, we read its probability of answering \texttt{null} rather than a PR number, averaged over the original and reversed PR order. Above a cut-off, the model answers ``no fix''; below it, the model keeps its answer from the main run. We compute this probability with the reason written before the answer, as in the main runs (\emph{reason first}), and with the answer field first (\emph{answer first}). We tune the cut-off on development issues and repeat the analysis with Qwen3-14B.

\subsection{Models and Baselines}
We evaluate eleven LLM setups: eight open models from the Qwen, Phi, and Granite families (1.7B--14B, including Qwen3-4B in thinking mode) and three hosted models run through an API (Gemini~3 Flash, the open-weight gpt-oss-120b, and Claude Sonnet~5). Qwen3-4B-Instruct-2507 was selected as the main open model on development issues. Open models run with vLLM~0.32 on two T4 GPUs at temperature~0, seed~0, a 16,384-token context, and up to 512 output tokens (6,144 for thinking mode). Hosted models use temperature~0 where supported and their default reasoning; Sonnet used no extended thinking. All models get the same truncated inputs: issue body up to 2,500 characters, comments up to the first 1,000 and last 1,500, PR description up to 1,200, and review discussion up to 600. Gemini~3 Flash and gpt-oss-120b also run with the strict prompt. On the 536 issues with a merged PR, Sonnet and gpt-oss-120b are also tested with fix claims removed from the evidence list and with the issue timeline, commits, and diffs provided.

We compare these models with six non-LLM methods, all restricted to merged PRs:
\begin{itemize}
    \item \textbf{Keyword rule:} selects a merged PR that says ``fixes,'' ``closes,'' or ``resolves'' the issue; otherwise ``no fix.''
    \item \textbf{Single merged PR rule:} selects the PR if exactly one is merged; otherwise ``no fix.''
    \item \textbf{Merge-first rule:} selects a merged PR whenever one exists.
    \item \textbf{Logistic regression:} scores PRs using metadata.
    \item \textbf{Embedding ranker} and \textbf{reranker} (bge-reranker-v2-m3): score the textual match between an issue and each merged PR.
\end{itemize}
The regression and the two rankers answer ``no fix'' when no PR scores above a cut-off tuned on development issues.

\subsection{Measures}
\label{sec:measures}
We use the 305 development issues for model selection, regression training, and cut-off tuning, and evaluate on the 675 test issues from separate repositories. The main prompt, answer reader, and model settings were fixed before testing. An answer is correct if it selects the labeled fix, or answers ``no fix'' when none exists.

For RQ3, we report accuracy separately on the 476 issues with a merged fix and the 60 with merged PRs but no fix, along with \emph{balanced accuracy}, the average of these two rates. Because we oversampled the rare group, balanced accuracy measures behavior on each group independently of our sample mix; Section~\ref{sec:discussion} weights the groups by their estimated frequency instead. Cut-offs are tuned for balanced accuracy. We also report accuracy on the 147 issues with several merged PRs and overall accuracy on all 675 test issues. We compute bootstrap 95\% confidence intervals (${\geq}1{,}000$ samples) and use exact McNemar tests, with Bonferroni correction for model comparisons and Holm correction for prompt/input changes and PR-feature analyses.

\section{Results}\label{sec:results}
\providecommand{\rqanswer}[2]{%
  \begin{tcolorbox}[
    colback=black!4, colframe=black!55,
    boxrule=0.5pt, arc=1mm,
    left=4pt, right=4pt, top=2pt, bottom=2pt,
    fontupper=\small]
    \textbf{#1.} #2
  \end{tcolorbox}
}

\begin{table*}[t]
\centering
\begin{threeparttable}
\caption{Results on the 536 test issues with at least one merged PR (accuracy, \%).}
\label{tab:main_results}
\small
\setlength{\tabcolsep}{5pt}
\renewcommand{\arraystretch}{1.12}
\begin{tabular*}{0.98\textwidth}{@{\extracolsep{\fill}}lrrrrrr@{}}
\toprule
\textbf{Method} &
\shortstack{\textbf{Merged fix}\\\textbf{found}\\($n$=476)} &
\shortstack{\textbf{Correct}\\\textbf{``no fix''}\\($n$=60)} &
\shortstack{\textbf{Balanced}\\\textbf{acc.}\\\phantom{($n$)}} &
\shortstack{\textbf{2+ merged}\\\textbf{PRs}\\($n$=147)} &
\shortstack{\textbf{All 675}\\\textbf{issues}\\\phantom{($n$)}} &
\shortstack{\textbf{Cost per}\\\textbf{1k issues}\\\phantom{($n$)}} \\
\midrule
\rowcolor{black!8}
\multicolumn{7}{@{}l}{\textbf{Non-LLM methods}} \\
Keyword rule                   & 40.5 & 95.0 & 67.8 & 51.7 & 56.7 & -- \\
Single merged PR rule          & 71.4 & 18.3 & 44.9 & 7.5  & 71.7 & -- \\
Merge-first rule               & 83.4 & 0.0  & 41.7 & 38.8 & 78.5 & -- \\
Logistic regression            & 35.9 & 91.7 & 63.8 & 19.7 & 53.2 & -- \\
Reranker                       & 68.7 & 55.0 & 61.8 & 53.7 & 73.0 & -- \\
Embedding ranker               & 61.3 & 81.7 & 71.5 & 56.5 & 70.2 & -- \\
\addlinespace[3pt]
\midrule
\rowcolor{black!8}
\multicolumn{7}{@{}l}{\textbf{Open models (two T4 GPUs)}} \\
Qwen3-1.7B                    & 89.7 & 28.3 & 59.0 & 70.1 & 85.5 & 8 min \\
Phi-4-mini (3.8B)             & 87.8 & 43.3 & 65.6 & 67.3 & 85.5 & 12 min \\
Granite-3.3-8B                & 95.2 & 13.3 & 54.3 & 77.6 & 88.0 & 28 min \\
Qwen3.5-4B                    & 92.9 & 58.3 & 75.6 & 78.9 & 90.4 & 13 min \\
Qwen3-8B                      & 93.3 & 58.3 & 75.8 & 79.6 & 90.7 & 23 min \\
\textbf{Qwen3-4B-Instruct (main)} & \textbf{96.0} & \textbf{53.3} & \textbf{74.7} & \textbf{83.0} & \textbf{92.1} & \textbf{21 min} \\
Qwen3-14B-AWQ                 & 95.4 & 65.0 & 80.2 & 85.0 & 92.7 & 35 min \\
Qwen3-4B (thinking mode)      & 95.2 & 55.0 & 75.1 & 81.6 & 91.7 & 4.6 h \\
\addlinespace[3pt]
\midrule
\rowcolor{black!8}
\multicolumn{7}{@{}l}{\textbf{API models}} \\
Gemini 3 Flash                & 95.6 & 61.7 & 78.6 & 84.4 & 92.6 & \$1.05 \\
gpt-oss-120b (5.1B active)    & 94.1 & 75.0 & 84.6 & 83.7 & 92.7 & \$0.23 \\
Claude Sonnet 5               & 93.1 & 88.3 & 90.7 & 83.7 & 93.2 & \$5.36 \\
\bottomrule
\end{tabular*}
\begin{tablenotes}[flushleft]
\footnotesize
\item All answers after the merged-only rule (Section~\ref{sec:mergedonly}). Balanced acc.: average of the first two columns. All 675 issues: includes the 139 with no merged PR, where every method is right on 133. Regression and ranker cut-offs were tuned for balanced accuracy on development issues; tuned for overall accuracy instead, they score 90.8/0.0 (regression), 85.5/20.0 (reranker), and 88.9/15.0 (embedding) on the first two columns. Cost: run time on two T4 GPUs (open models) or billed dollars (API).
\end{tablenotes}
\end{threeparttable}
\end{table*}

\begin{table}[t]
\centering
\caption{Unmerged picks on the 675 test issues (original answers, before the merged-only rule).}
\label{tab:unmerged}
\footnotesize
\setlength{\tabcolsep}{4pt}
\renewcommand{\arraystretch}{1.1}
\begin{tabular*}{\columnwidth}{@{\extracolsep{\fill}}lrr@{}}
\toprule
\textbf{Model} &
\shortstack{\textbf{Base prompt}\\\textbf{(with exception)}} &
\shortstack{\textbf{Strict prompt}\\\textbf{(no exception)}} \\
\midrule
Qwen3-1.7B                    & 151 & 140 \\
Phi-4-mini (3.8B)             & 77  & 35 \\
Granite-3.3-8B                & 137 & 102 \\
Qwen3.5-4B                    & 56  & 10 \\
Qwen3-8B                      & 109 & 39 \\
\textbf{Qwen3-4B-Instruct (main)} & \textbf{83} & \textbf{13} \\
Qwen3-14B-AWQ                 & 56  & 1 \\
Qwen3-4B (thinking mode)      & 7   & -- \\
\addlinespace[2pt]
Gemini 3 Flash                & 27  & 0 \\
gpt-oss-120b                  & 3   & 0 \\
Claude Sonnet 5               & 1   & -- \\
\bottomrule
\end{tabular*}
\vspace{2pt}
\par\footnotesize\raggedright
\emph{Note:} Unmerged picks: wrong picks of an unmerged PR, or of a PR not in the list (0--5 per model). --: not run. Base prompt: the main runs of Table~\ref{tab:main_results}.
\par
\end{table}

\begin{table}[t]
\centering
\caption{Changing the prompt or input.}
\label{tab:variants}
\footnotesize
\setlength{\tabcolsep}{3pt}
\renewcommand{\arraystretch}{1.1}
\begin{tabularx}{\columnwidth}{@{}>{\raggedright\arraybackslash}Xcccccc@{}}
\toprule
& \multicolumn{2}{c}{\textbf{4B acc.\ (\%)}} &
\multicolumn{4}{c}{\textbf{Unmerged picks}} \\
\cmidrule(lr){2-3}\cmidrule(l){4-7}
\textbf{Change} & \textbf{Orig.} & \textbf{Filt.} &
\textbf{4B} & \textbf{14B} & \textbf{Phi} & \textbf{Gran.} \\
\midrule
Base prompt (Fig.~\ref{fig:prompt}) & 80.9 & 92.1 & 83 & 56 & 75 & 137 \\
\addlinespace[2pt]
\rowcolor{black!8}
\multicolumn{7}{@{}l}{\textit{Changing the prompt}} \\
\quad + stricter rule & 81.6 & 92.0 & 77 & 20 & -- & -- \\
\quad -- exception (strict prompt) & 90.1 & 91.6 & 13 & 1 & 35 & 102 \\
\quad + merge flag on each PR & 88.6 & 90.7 & 18 & 28 & -- & -- \\
\quad + ``no fix'' is valid & 82.4 & 91.9 & 70 & 36 & 73 & 129 \\
\quad -- fix claims as evidence & 83.1 & 91.9 & 65 & 50 & 75 & 132 \\
\addlinespace[2pt]
\rowcolor{black!8}
\multicolumn{7}{@{}l}{\textit{Changing the input}} \\
\quad + timeline, commits, diffs & 79.9 & 92.0 & 89 & -- & -- & -- \\
\quad -- fix claims in unmerged PRs & 82.5 & 92.1 & 71 & 30 & 69 & 132 \\
\quad Merge status hidden & 65.8 & 85.8 & 184 & -- & 153 & 185 \\
\quad Merge status flipped & 36.4 & 57.2 & 262 & -- & 269 & 267 \\
\quad Unmerged PRs removed & 91.7 & 91.9 & 1 & -- & -- & -- \\
\bottomrule
\end{tabularx}
\vspace{2pt}
\par\footnotesize\raggedright
\emph{Note:} 4B acc.: Qwen3-4B's accuracy on all 675 test issues, before (Orig.) and after (Filt.) the merged-only rule. Unmerged picks are counted on original answers. Flipping the merge status also removes merge times. With unmerged PRs removed, the one remaining pick names a PR not in the list. --: not run. Phi and Granite base rows are a same-session rerun (Phi: 75 here, 77 in Table~\ref{tab:unmerged}).
\par
\end{table}

\begin{table}[t]
\centering
\caption{PR features and how often models picked unmerged PRs (base prompt).}
\label{tab:neglect}
\footnotesize
\setlength{\tabcolsep}{3pt}
\renewcommand{\arraystretch}{1.12}
\begin{tabularx}{\columnwidth}{@{}>{\raggedright\arraybackslash}Xrrcc@{}}
\toprule
\textbf{Feature} & \textbf{With} & \textbf{W/o} &
\shortstack{\textbf{Adj.\ OR}\\\textbf{[95\% CI]}} & \textbf{Models} \\
\midrule
No merged PR          & 26.0 & 5.4  & 4.5 [2.6, 10.0] & 6 (7) \\
Says ``Fixes \#N''    & 22.6 & 10.8 & 2.9 [1.7, 5.4]  & 4 (7) \\
\addlinespace[2pt]
Largest change        & 22.6 & 11.4 & 1.2 [0.6, 2.3]  & 4 (7) \\
Fix-like title        & 17.9 & 15.6 & 0.9 [0.5, 1.6]  & 0 (6) \\
Still open            & 20.4 & 15.5 & 1.0 [0.5, 2.0]  & 0 (6) \\
Listed first          & 17.5 & 15.5 & 0.8 [0.4, 1.6]  & 1 (4) \\
\bottomrule
\end{tabularx}
\vspace{2pt}
\par\footnotesize\raggedright
\emph{Note:} 484 unmerged test PRs that are not the fix. With / W/o: how often Qwen3-4B picked the PR with and without the feature (\%). Adj.\ OR: Qwen3-4B's adjusted odds ratio, with a bootstrap 95\% CI over issues. Models: number of the seven open models with a significant link after Holm correction (in brackets: with a positive link).
\par
\end{table}

\subsection{RQ1: How Often Do Closing Keywords Point to the Fix?}
\label{sec:rq1}

On the 1,051 labeled issues, the strict keyword rule (a merged PR saying ``fixes,'' ``closes,'' or ``resolves'' the issue) was correct on 84.3\% of its picks and found 41.2\% of fixes. Of its 57 wrong picks, 41 were helper PRs that claimed a fix but made only related changes. A looser rule picking merged PRs that mention the issue number found more fixes (91.5\%) but was right less often (71.9\%). Weighting by sampling group to estimate performance across all mined issues, the strict rule was correct on 96.1\% of its picks (95\% CI: 93.9--97.9) and found 64.7\% of fixes (60.1--69.2).

The opposite simple rule fails in the other direction. Picking any merged PR (the merge-first rule) found 83.4\% of merged fixes on the test set, but answered ``no fix'' correctly on none of the 60 issues whose merged PRs did not fix the problem (Table~\ref{tab:main_results}).

\rqanswer{Answer to RQ1}{Closing keywords are a precise but incomplete signal, and their mistakes are mostly merged helper PRs. Trusting merge status alone fails the other way: it never recognizes that no PR fixed the issue.}

\subsection{RQ2: Do LLMs Follow the Rule That Unmerged PRs Are Not Fixes?}
\label{sec:rq2}

We treat RQ2 as a case study of how models over-apply a rarely needed exception. We count \emph{unmerged picks} on the original answers, before the merged-only rule: wrong picks of an unmerged PR or of a PR not in the candidate list.

\noindent\textbf{A rare exception.}
The base prompt allowed an unmerged PR when a maintainer stated that its changes landed another way. This exception was needed for only 6 of 482 test fixes (1.2\%). Yet allowing it sharply increased unmerged picks for five of the seven open models (Table~\ref{tab:unmerged}): Qwen3-4B rose from 13 to 83, Qwen3-14B from 1 to 56, and Qwen3-8B from 39 to 109. Gemini~3 Flash rose from 0 to 27, while gpt-oss-120b rose only from 0 to 3.

\noindent\textbf{The strict prompt.}
Without the exception, most models largely followed the rule. Gemini~3 Flash and gpt-oss-120b made no unmerged picks, Qwen3-14B made one, and Qwen3-4B made 13. Granite-3.3-8B and Qwen3-1.7B were the exceptions, with 102 and 140 unmerged picks. These two models also rarely answered ``no fix,'' even though 28.6\% of test issues have no labeled fix.

\noindent\textbf{Merge status and fix claims.}
Models used merge status as the prompt instructed. Hiding it increased Qwen3-4B's unmerged picks from 83 to 184, and flipping it increased them to 262 (Table~\ref{tab:variants}). ``Fixes~\#N'' claims also mattered: removing them from unmerged PRs reduced unmerged picks for Qwen3-14B from 56 to 30 and for Qwen3-4B from 83 to 71. After the merged-only rule, Qwen3-4B's accuracy stayed between 90.7\% and 92.1\% across prompt changes, so these changes affected the kind of mistakes more than final accuracy.

\noindent\textbf{When unmerged picks happen.}
Among the 484 unmerged test PRs that are not the fix (Table~\ref{tab:neglect}), Qwen3-4B picked them more often when the issue had no merged PR (26.0\% vs.\ 5.4\%; adjusted odds ratio 4.5) and when the PR said ``Fixes~\#N'' (22.6\% vs.\ 10.8\%; odds ratio 2.9). List position and open status showed no clear link for Qwen3-4B. For four of the seven open models, picks were also more common when the PR had the largest change.

\rqanswer{Answer to RQ2}{Mostly, but several models over-applied a single rarely needed exception. The merged-only rule absorbed this failure, so such rules are best enforced in code rather than in the prompt.}

\subsection{RQ3: Can LLMs Tell When a Merged PR Did Not Fix the Issue?}
\label{sec:rq3}

Table~\ref{tab:main_results} compares all methods on the 536 test issues with at least one merged PR, after the merged-only rule. Table~\ref{tab:top5} extracts the five setups with the highest overall accuracy.

\begin{table}[t]
\centering
\caption{The five setups with the highest overall accuracy (\%, from Table~\ref{tab:main_results}).}
\label{tab:top5}
\footnotesize
\setlength{\tabcolsep}{4pt}
\renewcommand{\arraystretch}{1.1}
\begin{tabular*}{\columnwidth}{@{\extracolsep{\fill}}lrrr@{}}
\toprule
\textbf{Model} &
\shortstack{\textbf{Merged fix}\\\textbf{found} (476)} &
\shortstack{\textbf{Correct}\\\textbf{``no fix'' (60)}} &
\shortstack{\textbf{All}\\\textbf{675}} \\
\midrule
Qwen3-4B        & 96.0 & 53.3 & 92.1 \\
Gemini 3 Flash  & 95.6 & 61.7 & 92.6 \\
Qwen3-14B       & 95.4 & 65.0 & 92.7 \\
gpt-oss-120b    & 94.1 & 75.0 & 92.7 \\
Claude Sonnet 5 & 93.1 & 88.3 & 93.2 \\
\midrule
\textbf{Range}  & \textbf{2.9} & \textbf{35.0} & \textbf{1.1} \\
\bottomrule
\end{tabular*}
\end{table}

\noindent\textbf{Finding merged fixes is similar across models.}
The five setups found 93.1--96.0\% of merged fixes. Among the 476 issues with a merged fix, Sonnet and Qwen3-4B picked a wrong merged PR 17 and 16 times, respectively. On the 147 issues with several merged PRs, Qwen3-4B was correct on 83.0\% and the hosted models on 83.7--84.4\%.

\noindent\textbf{Noticing ``no fix'' is not.}
On the 60 issues where PRs were merged but none was the fix, the same five setups answered ``no fix'' correctly on 53.3--88.3\%. Sonnet was correct on 88.3\% (95\% CI: 80.0--96.7) and Qwen3-4B on 53.3\% (40.0--66.7); the paired difference was significant (24 vs.\ 3 discordant issues, $p<0.001$). Issue \#2617 from Section~\ref{sec:introduction} is one such case: Qwen3-4B picked the merged PR, while Sonnet answered ``no fix.'' Of Qwen3-4B's 32 correct answers, 10 came from the merged-only rule rather than the model's own answer; all 53 correct Sonnet answers were the model's own. The difference between Sonnet and each locally run model was significant after Bonferroni correction ($p\leq0.005$); the difference from gpt-oss-120b was not. A second Sonnet run was correct on 86.7\%.

Within the Qwen family, accuracy on this group grew with model size: 28.3\% (1.7B), 53.3--58.3\% (4B), 58.3\% (8B), and 65.0\% (14B), while finding merged fixes did not. Some non-LLM methods answered ``no fix'' correctly more often but found far fewer real fixes: the embedding ranker was correct on 81.7\% of these issues but found only 61.3\% of merged fixes.

\noindent\textbf{Answering ``no fix'' more often costs real fixes.}
On the 476 merged-fix issues, Sonnet wrongly answered ``no fix'' 16 times, compared with three times for Qwen3-4B (all three caused by the merged-only rule; 14 vs.\ 1 discordant issues, $p=0.001$).

\noindent\textbf{Prompt and input changes.}
Changing the prompt, including stating that ``no fix'' is valid and removing fix claims from the evidence list, left Qwen3-4B correct on 31--33 of the 60 merged-no-fix issues. Removing fix claims also left the other models near their base rates. Adding the issue timeline, commits, and diffs gave 60.0\% for Qwen3-4B, 81.7\% for gpt-oss-120b (from 75.0\%; $p=0.29$), and 85.0\% for Sonnet (from 88.3\%; $p=0.63$) on this group. Code thus helped some models a little, but Sonnet still led Qwen3-4B by 25 points, so missing diffs explain only a small part of the difference.

\begin{figure}[t]
  \centering
  \includegraphics[width=\columnwidth]{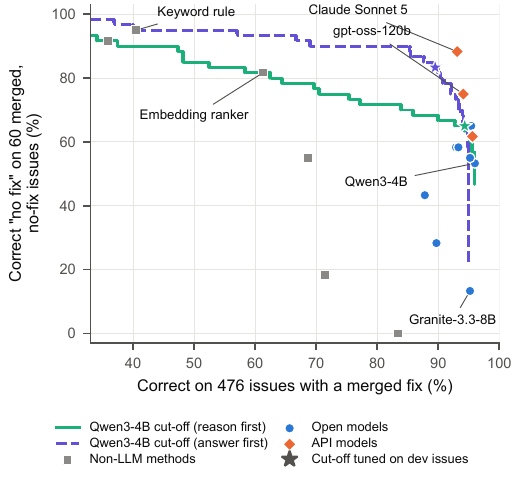}
  \caption{Finding merged fixes vs.\ answering ``no fix'' on the 536 test issues with a merged PR. Models with similar overall accuracy sit at different points of this trade-off. Points: Table~\ref{tab:main_results}. Lines: changing Qwen3-4B's cut-off trades found fixes for correct ``no fix'' answers. Stars: cut-offs tuned on development issues.}
  \label{fig:tradeoff}
\end{figure}

\noindent\textbf{Changing the cut-off.}
Changing Qwen3-4B's cut-off traded found fixes for correct ``no fix'' answers (Fig.~\ref{fig:tradeoff}), but Sonnet's point lies above both curves. At the cut-off matching Sonnet's 88.3\% on merged-no-fix issues, Qwen3-4B found 48.1\% of merged fixes with reason-first probabilities and 85.5\% with answer-first probabilities, and Qwen3-14B 76.7\% and 81.5\%, against Sonnet's 93.1\%. This comparison needs no development data. Both tuning choices favor the open models: the prompt was developed with Qwen3-4B, and only the open models get a tuned cut-off. With cut-offs tuned on development issues, the balanced-accuracy difference between Sonnet and Qwen3-4B was 11.0 points (95\% CI: 4.2--18.2) with reason first and 4.3 points ($-$1.5--9.9) with answer first. For Qwen3-14B, the differences were 6.0 points (1.1--11.6) and 4.6 points ($-$0.1--9.3). With answer-first probabilities, the remaining difference was not statistically significant for either open model.

\noindent\textbf{The difference holds across annotators.}
Each of the two additional annotators (Section~\ref{sec:labeling}) agreed with our labels on 90 of their 100 issues ($\kappa=0.81$), and they agreed with each other on 89\% ($\kappa=0.80$). Under our labels, either annotator's labels, or majority labels, Sonnet answered ``no fix'' correctly on 88--96\% of merged-no-fix issues and Qwen3-4B on 51--57\%.

\noindent\textbf{How common this case is.}
Our sample over-represents merged-no-fix issues (9.3\% of labeled issues). Weighting by sampling group, we estimate that they make up 3.9\% (95\% CI: 2.7--5.2) of the 7,765 mined issues outside the guideline and excluded sets, and 5.0\% (3.4--6.7) of those with a merged PR. Because the case is uncommon, overall accuracy hides it: all five setups in Table~\ref{tab:top5} scored 92--93\% overall.

\rqanswer{Answer to RQ3}{Sometimes. Models agree on finding merged fixes but differ widely in noticing when no merged PR is the fix, and this ability grows with model size. Sonnet's lead is not a threshold effect: no cut-off of the open models reached its point.}

\subsection{Other Observations}
\label{sec:other}

\noindent\textbf{Issue age.}
Some public test issues may have appeared in model training data. Splitting issues at August 2025 changed accuracy by at most 3.2 points, and every model performed at least as well on the 165 issues from 2026. We found no clear pattern by issue age.

\noindent\textbf{Labels for the remaining issues.}
We applied Qwen3-4B with the merged-only rule to the remaining 6,785 mined issues. It kept nearly all keyword links and found fixes for 1,469 issues without one. Such new links were correct 89.2\% of the time on labeled test issues. Weighted to the mined issues, we estimate that 97.6\% of the additional labels are correct (95\% CI: 95.8--98.9). In a separate sample of 100 issues, two annotators judged the labels correct on 93 of 96 and 92 of 94 in-scope issues, respectively. We release keyword-based and newly found links separately. Qwen3-4B is the top-five setup least likely to notice that no merged PR is the fix, so users who need fewer wrong references can re-run a more cautious model on the released prompts.

\section{Discussion and Threats to Validity}\label{sec:discussion}
\subsection{Implications}

\noindent\textbf{For building benchmarks.}
To see what the RQ3 differences mean for a benchmark, we apply the observed test error rates to 1,000 issues with a merged PR, of which an estimated 5.0\% have no merged fix (Section~\ref{sec:rq3}). Table~\ref{tab:projection} shows the result. These are illustrative projections, not measured pipeline outcomes.

\begin{table}[t]
\centering
\caption{Projected mistakes per 1,000 issues with a merged PR.}
\label{tab:projection}
\footnotesize
\setlength{\tabcolsep}{6pt}
\renewcommand{\arraystretch}{1.1}
\begin{tabular}{@{}lcc@{}}
\toprule
\textbf{Model} & \shortstack{\textbf{Wrong fixes}\\\textbf{kept}} & \shortstack{\textbf{Real fixes}\\\textbf{missed}} \\
\midrule
Qwen3-4B        & 55 & 6 \\
Qwen3-14B       & 45 & 16 \\
Claude Sonnet 5 & 40 & 32 \\
\bottomrule
\end{tabular}
\end{table}

Choosing a model changes the kind of mistakes more than their number. Qwen3-4B makes fewer mistakes if both kinds count equally, while Sonnet keeps about a quarter fewer wrong fixes: 95.7\% of the fixes Sonnet keeps would be correct, against 94.3\% for Qwen3-4B. Execution filters such as fail-to-pass tests do not remove these wrong fixes, since they check that a PR's own tests pass, not that it fixes the issue. If a wrong fix is worse than a dropped issue, a model that answers ``no fix'' more often is preferable; if task volume matters more, the reverse holds. Whichever model is used, keywords alone should not decide the fix, and the merge rule belongs in code rather than in the prompt.

\noindent\textbf{For evaluating models.}
We recommend reporting accuracy separately for issues with a merged fix and for issues whose merged PRs did not fix the problem, alongside overall accuracy. This separates failing to find a real fix from failing to notice that no PR is the fix.

\noindent\textbf{For future work.}
A natural next step is to apply this analysis to existing benchmarks. Audits of curated benchmarks show that reference links do go wrong (Section~\ref{sec:related}), and automated pipelines that pick a reference PR from several linked PRs without human review~\cite{badertdinov2025swerebench,zhang2025swebenchlive} are where our findings apply most directly; measuring how many of their reference fixes would change is left to future work. Other directions include comparing against issue--PR pair checkers~\cite{wang2026paichecker} and link-recovery tools adapted to choosing among PRs, training models specifically for picking the fix, adding executed tests as evidence, and sending uncertain cases to larger models. Our issue groups provide a way to evaluate such approaches separately.

\subsection{Threats to Validity}
\label{sec:threats}

\noindent\textbf{Merge status.}
Merging is strong evidence of acceptance, but fixes can also land through separate commits. The merged-only rule turned three correct Qwen3-4B answers into mistakes, and our check for silent landings (Section~\ref{sec:mergerate}) cannot rule out landings that leave no observable signal. The rule may exclude more real fixes in projects where changes often land without merging PRs.

\noindent\textbf{Labels based on records.}
Our labels rely on saved issue discussions and PR records rather than code diffs or executed tests. They identify the fix supported by the available records, which is also all an automated pipeline can see, not a behaviorally verified fix, and labels informed by code inspection or execution could differ, although adding diffs barely changed Sonnet's answers on merged-no-fix issues (Section~\ref{sec:rq3}). Our task also depends on the PRs recovered during mining, and we excluded issues with several equally important fixes.

\noindent\textbf{Label agreement.}
One author labeled all issues, which may introduce a consistent reading of what counts as the fix. The second annotator's agreement (Section~\ref{sec:labeling}) and the stability of the RQ3 difference under the two additional annotators' labels (Section~\ref{sec:rq3}) reduce this risk, although agreement was lower on the RQ3 sample.

\noindent\textbf{Sampling and group size.}
We oversampled unclear issues and competing PRs. Our weighted estimates assume that labeled issues represent their sampling groups, and they apply to the mined issues rather than all GitHub issues. The merged-no-fix test group contains only 60 issues, which gives wide confidence intervals and limits fine comparisons between models with similar rates; the intervals for Sonnet and Qwen3-4B, however, do not overlap. The projections in Table~\ref{tab:projection} also assume that the observed error rates per group apply to the estimated issue mix.

\noindent\textbf{Models and prompts.}
Our findings are limited to the tested models, prompts, and inference settings. We tested one wording of the unmerged-PR exception; other wordings may change model behavior. Sonnet was not tested with the strict prompt because it picked an unmerged PR only once under the base prompt. Main setups were generally run once at temperature~0, so repeated API runs may differ. We changed the ``no fix'' cut-off only for Qwen3-4B and Qwen3-14B. Public test issues may have appeared in model training data; the issue-age analysis (Section~\ref{sec:other}) reduces but does not remove this concern.
\section{Conclusion}\label{sec:conclusion}
Picking the fix for a GitHub issue looks easy when a merged PR says ``Fixes \#N,'' and our 1,051 labeled issues show where that intuition breaks. Keywords and merge status are useful signals but cannot decide the fix on their own, and rules such as the merge rule are safer in code than in a prompt. LLMs agree on most issues; where they part ways is the uncommon case in which work was merged but nothing was fixed. Catching that case costs real fixes, so building a benchmark from issues means choosing which mistakes to accept, and evaluating these methods means looking past a single accuracy number.

\section*{Data Availability}
The dataset, prompts, model outputs, analysis scripts, and additional labels produced by a model will be made publicly available upon acceptance.

\bibliographystyle{IEEEtran}
\bibliography{ref}

@inproceedings{bird2009bias,
  author    = {Bird, Christian and Bachmann, Adrian and Aune, Eirik and Duffy, John and Bernstein, Abraham and Filkov, Vladimir and Devanbu, Premkumar},
  title     = {Fair and Balanced? Bias in Bug-Fix Datasets},
  booktitle = {Proceedings of the Joint Meeting of the European Software Engineering Conference and the ACM SIGSOFT Symposium on the Foundations of Software Engineering},
  pages     = {121--130},
  year      = {2009},
  publisher = {ACM},
  doi       = {10.1145/1595696.1595716}
}

@inproceedings{wu2011relink,
  author    = {Wu, Rongxin and Zhang, Hongyu and Kim, Sunghun and Cheung, Shing-Chi},
  title     = {{ReLink}: Recovering Links Between Bugs and Changes},
  booktitle = {Proceedings of the ACM SIGSOFT Symposium and European Conference on Foundations of Software Engineering},
  pages     = {15--25},
  year      = {2011},
  publisher = {ACM},
  doi       = {10.1145/2025113.2025120}
}

@article{sun2017frlink,
  author  = {Sun, Yan and Wang, Qing and Yang, Ye},
  title   = {{FRLink}: Improving the Recovery of Missing Issue-Commit Links by Revisiting File Relevance},
  journal = {Information and Software Technology},
  volume  = {84},
  pages   = {33--47},
  year    = {2017},
  doi     = {10.1016/j.infsof.2016.11.010}
}

@article{ruan2019deeplink,
  author  = {Ruan, Hang and Chen, Bihuan and Peng, Xin and Zhao, Wenyun},
  title   = {{DeepLink}: Recovering Issue-Commit Links Based on Deep Learning},
  journal = {Journal of Systems and Software},
  volume  = {158},
  pages   = {110406},
  year    = {2019},
  doi     = {10.1016/j.jss.2019.110406}
}

@inproceedings{rostami2021hybridlinker,
  author    = {Rostami Mazrae, Pooya and Izadi, Maliheh and Heydarnoori, Abbas},
  title     = {Automated Recovery of Issue-Commit Links Leveraging Both Textual and Non-textual Data},
  booktitle = {Proceedings of the IEEE International Conference on Software Maintenance and Evolution (ICSME)},
  pages     = {263--273},
  year      = {2021},
  publisher = {IEEE},
  doi       = {10.1109/ICSME52107.2021.00030}
}

@inproceedings{zhang2023ealink,
  author    = {Zhang, Chenyuan and Wang, Yanlin and Wei, Zhao and Xu, Yong and Wang, Juhong and Li, Hui and Ji, Rongrong},
  title     = {{EALink}: An Efficient and Accurate Pre-trained Framework for Issue-Commit Link Recovery},
  booktitle = {Proceedings of the 38th IEEE/ACM International Conference on Automated Software Engineering (ASE)},
  pages     = {217--229},
  year      = {2023},
  publisher = {IEEE},
  doi       = {10.1109/ASE56229.2023.00059}
}

@article{wang2025mplinker,
  author  = {Wang, Bangchao and Deng, Yang and Luo, Ruiqi and Liang, Peng and Bi, Tingting},
  title   = {{MPLinker}: Multi-template Prompt-tuning with Adversarial Training for Issue--Commit Link Recovery},
  journal = {Journal of Systems and Software},
  volume  = {223},
  pages   = {112351},
  year    = {2025},
  doi     = {10.1016/j.jss.2025.112351}
}

@inproceedings{huang2026easylink,
  author    = {Huang, Huihui and Widyasari, Ratnadira and Zhang, Ting and Irsan, Ivana Clairine and Shi, Jieke and Ang, Han Wei and Liauw, Frank and Ouh, Eng Lieh and Shar, Lwin Khin and Kang, Hong Jin and Lo, David},
  title     = {Back to the Basics: Rethinking Issue-Commit Linking with {LLM}-Assisted Retrieval},
  booktitle = {Proceedings of the 48th International Conference on Software Engineering (ICSE)},
  pages     = {2161--2173},
  year      = {2026},
  publisher = {ACM},
  doi       = {10.1145/3744916.3764530}
}

@article{zheng2026selink,
  author  = {Zheng, Tao and Zhao, Yangyang and Guo, Jiamin and Nie, Liming and Jiang, Mingyue and Zhou, Yuming},
  title   = {{SELink}: A Semantic-Enhanced Modular Framework for Issue--Commit Link Recovery},
  journal = {Information and Software Technology},
  volume  = {190},
  pages   = {107973},
  year    = {2026},
  doi     = {10.1016/j.infsof.2025.107973}
}

@article{akhavan2026linkanchor,
  author  = {Akhavan, Arshia and Hoseinpour, Alireza and Heydarnoori, Abbas and Bagheri, Hamid and Keshani, Mehdi},
  title   = {{LinkAnchor}: An Autonomous {LLM}-Based Agent for Issue-to-Commit Link Recovery},
  journal = {Proceedings of the ACM on Software Engineering},
  volume  = {3},
  number  = {FSE},
  pages   = {4184--4207},
  year    = {2026},
  doi     = {10.1145/3808191}
}

@inproceedings{morgan2026rethinking,
  author    = {Morgan, Cole and Asaduzzaman, Muhammad and Chowdhury, Shaiful and Wang, Shaowei},
  title     = {Think Harder and Don't Overlook Your Options: Revisiting Issue-Commit Linking with {LLM}-Assisted Retrieval},
  booktitle = {Proceedings of the International Conference on Evaluation and Assessment in Software Engineering (EASE)},
  year      = {2026},
  publisher = {ACM}
}

@inproceedings{isik2025prreviews,
  author    = {I{\c{s}}{\i}k, Ali Tunahan and {\c{C}}a{\u{g}}lar, Hatice K{\"u}bra and T{\"u}z{\"u}n, Eray},
  title     = {Enhancing Pull Request Reviews: Leveraging Large Language Models to Detect Inconsistencies Between Issues and Pull Requests},
  booktitle = {Proceedings of the IEEE/ACM Second International Conference on AI Foundation Models and Software Engineering (FORGE)},
  pages     = {168--178},
  year      = {2025},
  publisher = {IEEE},
  doi       = {10.1109/FORGE66646.2025.00027}
}

@inproceedings{wang2026paichecker,
  author    = {Wang, Manyi and Xu, Junjielong and He, Pinjia},
  title     = {{PAIChecker}: Uncovering and Checking {PR}-Issue Misalignment in {SWE}-Bench-Like Benchmarks},
  booktitle = {Proceedings of the 41st International Conference on Automated Software Engineering (ASE)},
  pages     = {3315--3327},
  year      = {2026},
  publisher = {ACM},
  doi       = {10.1145/3832783.3837557}
}

@inproceedings{jimenez2024swebench,
  author    = {Jimenez, Carlos E. and Yang, John and Wettig, Alexander and Yao, Shunyu and Pei, Kexin and Press, Ofir and Narasimhan, Karthik R.},
  title     = {{SWE}-bench: Can Language Models Resolve Real-world {GitHub} Issues?},
  booktitle = {Proceedings of the International Conference on Learning Representations (ICLR)},
  year      = {2024},
  url       = {https://openreview.net/forum?id=VTF8yNQM66}
}

@misc{openai2024swebenchverified,
  author       = {{OpenAI}},
  title        = {Introducing {SWE}-bench Verified},
  year         = {2024},
  howpublished = {OpenAI research article},
  month        = aug,
  url          = {https://openai.com/index/introducing-swe-bench-verified/}
}

@inproceedings{pan2025swegym,
  author    = {Pan, Jiayi and Wang, Xingyao and Neubig, Graham and Jaitly, Navdeep and Ji, Heng and Suhr, Alane and Zhang, Yizhe},
  title     = {Training Software Engineering Agents and Verifiers with {SWE}-Gym},
  booktitle = {Proceedings of the International Conference on Machine Learning (ICML)},
  series    = {Proceedings of Machine Learning Research},
  volume    = {267},
  pages     = {47717--47737},
  year      = {2025},
  publisher = {PMLR},
  url       = {https://proceedings.mlr.press/v267/pan25g.html}
}

@inproceedings{zhang2025swebenchlive,
  author    = {Zhang, Linghao and He, Shilin and Zhang, Chaoyun and Kang, Yu and Li, Bowen and Xie, Chengxing and Wang, Junhao and Wang, Maoquan and Huang, Yufan and Fu, Shengyu and Nallipogu, Elsie and Lin, Qingwei and Dang, Yingnong and Rajmohan, Saravan and Zhang, Dongmei},
  title     = {{SWE}-bench Goes Live!},
  booktitle = {Advances in Neural Information Processing Systems (NeurIPS), Datasets and Benchmarks Track},
  volume    = {38},
  pages     = {163347--163372},
  year      = {2025},
  doi       = {10.52202/085713-4923}
}

@inproceedings{badertdinov2025swerebench,
  author    = {Badertdinov, Ibragim and Golubev, Alexander and Nekrashevich, Maksim and Shevtsov, Anton and Karasik, Simon and Andriushchenko, Andrei and Trofimova, Maria and Litvintseva, Daria and Yangel, Boris},
  title     = {{SWE}-rebench: An Automated Pipeline for Task Collection and Decontaminated Evaluation of Software Engineering Agents},
  booktitle = {Advances in Neural Information Processing Systems (NeurIPS), Datasets and Benchmarks Track},
  volume    = {38},
  pages     = {26420--26466},
  year      = {2025},
  doi       = {10.52202/085713-0788}
}

@inproceedings{bhatia2025spice,
  author    = {Oliva, Gustavo A. and Rajbahadur, Gopi Krishnan and Bhatia, Aaditya and Zhang, Haoxiang and Chen, Yihao and Chen, Zhilong and Leung, Arthur and Lin, Dayi and Chen, Boyuan and Hassan, Ahmed E.},
  title     = {{SPICE}: An Automated {SWE}-Bench Labeling Pipeline for Issue Clarity, Test Coverage, and Effort Estimation},
  booktitle = {Proceedings of the 40th IEEE/ACM International Conference on Automated Software Engineering (ASE)},
  pages     = {2325--2337},
  year      = {2025},
  publisher = {IEEE},
  doi       = {10.1109/ASE63991.2025.00192}
}

@inproceedings{bachmann2010missing,
  author    = {Bachmann, Adrian and Bird, Christian and Rahman, Foyzur and Devanbu, Premkumar and Bernstein, Abraham},
  title     = {The Missing Links: Bugs and Bug-Fix Commits},
  booktitle = {Proceedings of the 18th ACM SIGSOFT International Symposium on Foundations of Software Engineering (FSE)},
  pages     = {97--106},
  year      = {2010},
  publisher = {ACM},
  doi       = {10.1145/1882291.1882308}
}

@inproceedings{xue2026swebenchplus,
  author    = {Xue, Haoran and Aleithan, Reem and Enan, Nafid and Uddin, Gias and Wang, Song},
  title     = {{SWE}-Bench+: Enhanced {LLM} Coding Benchmark},
  booktitle = {Proceedings of the 3rd ACM International Conference on AI-Powered Software (AIware)},
  pages     = {332--339},
  year      = {2026},
  publisher = {ACM},
  doi       = {10.1145/3805760.3814924}
}

@article{zhou2023ifeval,
  title   = {Instruction-Following Evaluation for Large Language Models},
  author  = {Zhou, Jeffrey and Lu, Tianjian and Mishra, Swaroop and Brahma, Siddhartha and Basu, Sujoy and Luan, Yi and Zhou, Denny and Hou, Le},
  journal = {arXiv preprint arXiv:2311.07911},
  year    = {2023}
}

@inproceedings{jiang2024followbench,
  title     = {{FollowBench}: A Multi-level Fine-grained Constraints Following Benchmark for Large Language Models},
  author    = {Jiang, Yuxin and Wang, Yufei and Zeng, Xingshan and Zhong, Wanjun and Li, Liangyou and Mi, Fei and Shang, Lifeng and Jiang, Xin and Liu, Qun and Wang, Wei},
  booktitle = {Proceedings of the 62nd Annual Meeting of the Association for Computational Linguistics (ACL)},
  year      = {2024}
}

@article{wen2025abstention,
  title   = {Know Your Limits: A Survey of Abstention in Large Language Models},
  author  = {Wen, Bingbing and Yao, Jihan and Feng, Shangbin and Xu, Chenjun and Tsvetkov, Yulia and Howe, Bill and Wang, Lucy Lu},
  journal = {Transactions of the Association for Computational Linguistics},
  volume  = {13},
  pages   = {529--556},
  year    = {2025},
  doi     = {10.1162/tacl_a_00754}
}

@inproceedings{kirichenko2025abstentionbench,
  title     = {{AbstentionBench}: Reasoning {LLMs} Fail on Unanswerable Questions},
  author    = {Kirichenko, Polina and Ibrahim, Mark and Chaudhuri, Kamalika and Bell, Samuel J.},
  booktitle = {Advances in Neural Information Processing Systems (NeurIPS)},
  year      = {2025}
}
\end{document}